\documentclass[final,10pt,technote,letterpaper]{IEEEtran}

\IEEEoverridecommandlockouts

\usepackage{amsmath}
\usepackage{amssymb}
\usepackage{hyperref}
\usepackage{array}
\usepackage{graphicx}
\usepackage[sectionbib]{bibunits}
\usepackage{tikz}
\usepackage{tkz-graph}
\usepackage[compress]{cite}
\usepackage{todonotes}
\usepackage{algorithmic}
\usepackage{algorithm}
\usepackage{xcolor}
\usepackage{enumitem}

\newcommand{\update}[1]{{\color{black}#1}}

\newtheorem{lemma}{Lemma}
\newtheorem{theorem}[lemma]{Theorem}
\newtheorem{definition}{Definition}
\newtheorem{remark}{Remark}
\newtheorem{assumption}{Assumption}

\DeclareMathOperator{\argmin}{arg\,min}
\DeclareMathOperator{\argmax}{arg\,max}

\let\oldbibliography\thebibliography
\renewcommand{\thebibliography}[1]{%
  \small
  \oldbibliography{#1}%
  \setlength{\itemsep}{0pt}%
}

\newcommand{\longpaper}{}

\PassOptionsToPackage{usenames,dvipsnames,svgnames}{xcolor}
\usetikzlibrary{arrows,positioning,automata,shapes,calc,shadows}

\hypersetup{
    colorlinks,
    citecolor=black,
    filecolor=black,
    linkcolor=black,
    urlcolor=black
}

\renewcommand{\matrix}[2]{\left[\begin{array}{#1}#2\end{array} \right]}

\title{ Mutual Information as Privacy-Loss Measure in Strategic Communication }

\author{Farhad Farokhi and Girish Nair\vspace{-.1in}\thanks{The authors are with the Department of Electrical and Electronic Engineering, The University of Melbourne, Parkville, Victoria 3010, Australia. Emails:\{ffarokhi,gnair\}@unimelb.edu.au}\thanks{The work of F.~Farokhi was supported by ARC grant LP130100605. The work of G. Nair was supported by ARC grants DP140100819 and FT140100527.}}

\begin{document}

\maketitle

\begin{abstract} A game is introduced to study the effect of privacy in strategic communication between well-informed senders and a receiver. The receiver wants to accurately estimate a random variable. The sender, however, wants to communicate a message that balances a trade-off between providing an accurate measurement and minimizing the amount of leaked private information, which is assumed to be correlated with the to-be-estimated variable. 
The mutual information between the transmitted message and the private information is used as a measure of the amount of leaked information. An equilibrium is constructed and its properties are investigated.
\end{abstract}

\section{Introduction}
Participatory and crowd-sensing technologies rely on honest data from recruited users to generate estimates of variables, such as traffic condition
and network coverage. 
However, providing accurate information by the users  undermines their privacy. For instance, a road user that provides her start and finish points as well as the travel time to a participatory-sensing scheme can significantly improve the quality of the traffic estimation; however, such reports expose her private life. Therefore, she benefits from providing ``false'' information, not to deceive the system and disrupt the services but to protect her privacy. The amount of the deviation from the truth is determined by the value of privacy, which varies across the population. To better understand this effect, here, we use a game-theoretic framework to model the conflict of interest and study the effect of privacy in strategic communication. 

Specifically, we use a model in which the receiver is interested in estimating a random variable. To this aim, the receiver ask a better-informed sender to provide a measurement of the variable. The sender wants to find a trade-off between her desire to provide an accurate measurement of the variable while minimizing the amount of leaked private information, which is potentially correlated with that variable. We assume that the sender has access to a possibly noisy measurement of the variable and a perfect measurement of her private information. We use the \textit{mutual information} between the communicated message and the private information to capture the amount of the leaked information. The sender balances between her two desires using a privacy ratio. We  present a numerical algorithm for finding an equilibrium (i.e., policies from which one has an incentive to unilaterally deviate) of the presented game. We also construct an equilibrium  explicitly for the case where the message and the to-be-estimated variable belong to the same space. Using a numerical example, we illustrate the relationship between the quality of the estimation at the receiver and the privacy ratio. In turns out that, at least for the presented example, there exists a critical value for the privacy ratio below which the sender honestly provides her measurement of the variable. 

Strategic communication has  been studied in the economics literature in the context of cheap-talk games~\cite{crawford1982strategic,farrell1996cheap,JobelSignalling} in which well-informed senders communicates with a receiver that makes a decision regarding the society's welfare. In those games, the sender(s) and the receiver have a clear conflict of interest, which results in potentially dishonest messages. Contrary to those studies, here, the conflict of interest is motivated by the sense of privacy of the sender, which changes the form of the cost functions. Furthermore, in this paper, we are dealing with discrete random variables which is different from the studies on cheap-talk games. Cheap-talk games were recently adapted to investigate privacy in communication and estimation~\cite{farokhi2014}. That study, however, focuses quadratic cost functions and Gaussian random variables. In this paper, we use the mutual information as a measure of the leaked information and study the more realistic setup of discrete communication channels. 

The problem considered in this paper is close, in essence, to the idea of differential privacy and its application in estimation and signal processing, e.g.~\cite{Dwork2008,friedman2010data,dwork2009differential}.
Those studies rely on adding noise, typically Laplace noises, to guarantee the privacy of the users by making the outcome less sensitive to local parameter variations. In contrast, here, we find the optimal ``amount of randomness'' that needs to be introduced into the system for preserving the privacy by modelling the communication as a strategic game and studying its equilibria.

In the information theory literature, wiretap channels have been studied heavily dating from the pioneering work in~\cite{6772207}. In these problems, the sender wishes to devise encoding schemes to create a secure channel for communicating with the receiver while hiding her data from an eavesdropper. 
However, in the privacy problem, the objective is different, that is, the sender want to hide her private information (not necessarily all the data possessed by her) from everyone including, but not limited to, the receiver. 

Note that information theory has been used in the past in networked control under communication constraints, e.g., see~\cite{tatikonda2004control,ishii2003limited, martins2008feedback,martins2007fundamental, baillieul2007control}. However, to the best of our knowledge, it has not been used to measure the privacy loss in strategic communication as in this paper. 

The rest of the paper is organized as follows. The  problem formulation is introduced in Section~\ref{sec:problem}. The equilibria of the communication game are constructed  in Section~\ref{sec:results}. The results are extended to the multi-sender case in Section~\ref{sec:multiple}. Section~\ref{sec:numerical} discusses the numerical example. The paper is concluded in Section~\ref{sec:conclusions}.

\begin{figure}[t]
\centering
\begin{tikzpicture}
\node[state,minimum size=0.1cm,scale=0.9] (R)  at (+1.0,+0.0) {$R$};
\node[state,minimum size=0.1cm,scale=0.9] (S)  at (-1.0,+0.0) {$S$};
\path[every node/.style={font=\sffamily\small}]
			(S)  		 edge [->,double=black] node[above] {$Y$} 		(R)
			(R)  		 edge [->,double=black] node[above] {$\hat{X}$} (+3.0,+0.0)
			(-3.0,+0.0)  edge [->,double=black] node[above] {$(Z,W)$} 	(S);
\end{tikzpicture}
\vspace{-.1in}
\caption{\label{fig:diagram0} Communication structure between the sender $S$ and the receiver $R$.}
\vspace*{-.2in}
\end{figure}
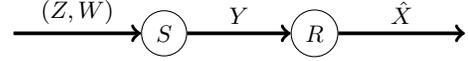

\section{Problem Formulation} \label{sec:problem}
We consider strategic communication between a sender  and a receiver as depicted in Fig.~\ref{fig:diagram0}. The receiver wants to have an accurate measurement of a discrete random variable $X\in\mathcal{X}$, where $\mathcal{X}$ denotes the set of all the possibilities. To this aim, the receiver deploys a sensor (which is a part of the sender) to provide a measurement of the variable. The measurement is denoted by $Z\in\mathcal{X}$. The sender also has another discrete random variable denoted by $W\in\mathcal{W}$, which is correlated with $X$ and/or $Z$. This random variable is the sender's private information, i.e., it is not known by the receiver. The sender wants to transmit a message $Y\in\mathcal{Y}$ that contains useful information about the measured variable while minimizing the amount of the leaked private information (note that, because of the correlation between $W$ and $X$ and/or $Z$, an honest report of $Z$ may shine some light on the realization of $W$).
Throughout this paper, for notational consistency, we use capital letters to denote the random variables, e.g., $X$, and small letters to denote a value, e.g., $x$. 

\begin{assumption} The discrete random variables $X,Z,W$ are distributed according to a joint probability distribution $p:\mathcal{X}\times\mathcal{X}\times \mathcal{W}\rightarrow [0,1]$, i.e., $\mathbb{P}\{X=x,Z=z,W=w\}=p(x,z,w)$ for all $(x,z,w)\in\mathcal{X}\times\mathcal{X}\times \mathcal{W}$.
\end{assumption}

The conflict of interest between the sender and the receiver can be modelled and analysed as a game. This conflict of interest can manifest itself in the following ways:
\begin{itemize}[leftmargin=*]
\item[1)] In participatory-sensing schemes, the sender's measurement of the state potentially depends on the way that the sender experiences the underlying process or services. For instance, in traffic estimation, the sender's measurement is fairly accurate on the route that she has travelled and, thus, an honest revelation of $Z$ provides a window into the life of the commuter. However, the underlying state $X$ is not related to the private information of sender $W$ since she is only an infinitesimal part of the traffic flow. In such a case, we have
\begin{align*}
\mathbb{P}&\{X=x,Z=z,W=w\}\\
&=\mathbb{P}\{Z=z|X=x,W=w\}\mathbb{P}\{X=x,W=w\}\\
&=\mathbb{P}\{Z=z|X=x,W=w\}\mathbb{P}\{X=x\}\mathbb{P}\{W=w\},
\end{align*}
where the second equality follows from independence of random variables $W$ and $X$.
\item[2)] In many services, such as buying insurance coverage or participating in polling surveys, an individual should provide an accurate history of her life or beliefs. In these cases, the variable $X$ highly depends on the private information of the sender $W$ (if not equal to it). In such cases,the measurement $Z$ may not contain any error as well. 
\end{itemize}
In what follows, the privacy game is properly defined.  

\subsection{Receiver} The receiver constructs its best estimate $\hat{X}\in\mathcal{X}$ using the conditional distribution
$\mathbb{P}\{\hat{X}=\hat{x}|Y=y\}=\beta_{\hat{x}y}$ for all $(\hat{x},y)\in\mathcal{X}\times\mathcal{Y}$. The matrix $\beta=(\beta_{\hat{x}y})_{(\hat{x},y)\in\mathcal{X}\times\mathcal{Y}}\in B$ is the policy of the receiver with the set of feasible policies defined~as
\begin{align*}
B\hspace{-.03in}=\hspace{-.03in}\bigg\{\hspace{-.03in}\beta\hspace{-.03in}:\hspace{-.03in}\beta_{\hat{x}y}\in[0,1],&\forall (\hat{x},y)\in\mathcal{X}\times\mathcal{Y}\,\&\hspace{-.03in}\sum_{\hat{x}\in\mathcal{X}}\hspace{-.03in}\beta_{\hat{x}y}\hspace{-.03in}=\hspace{-.03in}1,\forall y\in\mathcal{Y}\bigg\}.
\end{align*}
The receiver prefers an accurate measurement of the variable $X$. Therefore, the receiver wants to minimize the cost function $\mathbb{E}\{d(X,\hat{X})\}$ with the mapping $d:\mathcal{X}\times\mathcal{X}\rightarrow \mathbb{R}_{\geq 0}$ being a measure of distance between the entries of set $\mathcal{X}$. An example of such a distance is
\begin{align} \label{eqn:d}
d(x,\hat{x})=\left\{
\begin{array}{ll}
0, & x=\hat{x}, \\
1, & x\neq \hat{x}.
\end{array}
\right.
\end{align}
When using the distance mapping in~\eqref{eqn:d}, the term $\mathbb{E}\{d(X,\hat{X})\}$ becomes the probability of error at the receiver. The results of this paper are valid irrespective of the choice of this mapping.

\subsection{Sender} The sender constructs its message $y\in\mathcal{Y}$ according to the conditional probability distribution $\mathbb{P}\{Y=y|Z=z,W=w\}=\alpha_{yzw}$ for all $(y,z,w)\in\mathcal{Y}\times\mathcal{X}\times \mathcal{W}$. Therefore, the tensor $\alpha=(\alpha_{yzw})_{(y,z,w)\in\mathcal{Y}\times\mathcal{X}\times \mathcal{W}}\in A$ denotes the policy of the sender. The set of feasible policies is given by
\begin{align*}
A=\bigg\{\alpha: \,\alpha_{yzw}&\in[0,1],\forall (y,z,w)\in\mathcal{Y}\times\mathcal{X}\times\mathcal{W} \\[-0.7em]&\&\sum_{y\in\mathcal{Y}}\alpha_{yzw}=1,\forall (z,w)\in\mathcal{X}\times \mathcal{W}\bigg\}.
\end{align*}
The sender wants to minimize $\mathbb{E}\{d(X,\hat{X})\}+\varrho I(Y;W)$, with $I(Y;W)$ denoting the mutual information between random variables $Y$ and $W$~\cite{Elements2006}, to strike a balance between transmitting useful information about the measured variable and minimizing the amount of the leaked private information. In this setup, the privacy ratio $\varrho$ captures the sender's emphasis on protecting her privacy. For small $\varrho$, the sender provides a fairly honest measurement of the state. However, as $\varrho$ increases, the sender provides a less relevant message to avoid revealing her private information through the communicated message.

\subsection{Equilibria}
The cost function of the sender is equal
$$
U(\alpha,\beta)=\mathbb{E}\{d(X,\hat{X})\}+\varrho I(Y;W),
$$
where
\begin{align}
I(Y;W)=\sum_{y\in\mathcal{Y}}\sum_{w\in\mathcal{W}} &\mathbb{P}\{Y=y,W=w\} \nonumber \\& \times \log\left[\frac{\mathbb{P}\{Y=y,W=w\}}{\mathbb{P}\{Y=y\}\mathbb{P}\{W=w\}} \right]\label{eqn:mutual}
\end{align}
with $\mathbb{P}\{Y=y\}
=\sum_{w\in\mathcal{W}}\sum_{z\in\mathcal{X}}\sum_{x\in\mathcal{X}}\alpha_{yzw}p(x,z,w)$, $\mathbb{P}\{W=w\}=\sum_{z\in\mathcal{X}}\sum_{x\in\mathcal{X}}p(x,z,w)$, and
\begin{align*}
\mathbb{P}\{Y=y,W=w\}
&=\sum_{z\in\mathcal{X}}\mathbb{P}\{Y=y,W=w,Z=z\}\\
&=\sum_{z\in\mathcal{X}}\mathbb{P}\{Y=y|W=w,Z=z\}\\
&\hspace{.2in}\times\mathbb{P}\{W=w,Z=z\}\\
&=\sum_{z\in\mathcal{X}}\sum_{x\in\mathcal{X}}\alpha_{yzw}p(x,z,w).
\end{align*}
Moreover, we have
\begin{align*}
\mathbb{E}\{d(X,\hat{X})\}
&=\sum_{x\in\mathcal{X}}\sum_{\hat{x}\in\mathcal{X}}d(x,\hat{x})\mathbb{P}\{\hat{X}=\hat{x},X=x\}
\end{align*}
with
\begin{align*}
\mathbb{P}\{&\hat{X}=\hat{x},X=x\}
=\sum_{y\in\mathcal{Y}}\mathbb{P}\{\hat{X}=\hat{x},X=x,Y=y\}\\
&=\sum_{y\in\mathcal{Y}}\mathbb{P}\{\hat{X}=\hat{x}|X=x,Y=y\}\mathbb{P}\{X=x,Y=y\}\\
&=\sum_{y\in\mathcal{Y}}\beta_{\hat{x}y}\sum_{z\in\mathcal{X}}\sum_{w\in\mathcal{W}}\mathbb{P}\{Y=y|X=x,Z=z,W=w\}\\
&\hspace{1.4in}\times \mathbb{P}\{X=x,Z=z,W=w\}\\
&=\sum_{y\in\mathcal{Y}}\sum_{z\in\mathcal{X}}\sum_{w\in\mathcal{W}}\beta_{\hat{x}y}\alpha_{yzw}p(x,z,w).
\end{align*}
Following these calculations, we can define mappings $\xi:A\times B\rightarrow \mathbb{R}$ and $\zeta:A\rightarrow \mathbb{R}$ such that
\begin{align*}
\xi(\alpha,\beta)
&=\mathbb{E}\{d(X,\hat{X})\}\\
&=\sum_{x\in\mathcal{X}}\sum_{\hat{x}\in\mathcal{X}}\sum_{y\in\mathcal{Y}}\sum_{z\in\mathcal{X}}\sum_{w\in\mathcal{W}}d(x,\hat{x})\beta_{\hat{x}y}\alpha_{yzw}p(x,z,w)
\end{align*}
and
\begin{align*}
\zeta(\alpha)
&=I(Y;W)\\
&=\sum_{y\in\mathcal{Y}}\sum_{w\in\mathcal{W}} \bigg(\sum_{z\in\mathcal{X}}\sum_{x\in\mathcal{X}}\alpha_{yzw}p(x,z,w)\bigg) \nonumber \\& \times \log\left[\frac{\sum_{z\in\mathcal{X}}\sum_{x\in\mathcal{X}}\alpha_{yzw}p(x,z,w)}{(\sum_{w\in\mathcal{W},z\in\mathcal{X},x\in\mathcal{X}}\alpha_{yzw}p(x,z,w))\mathbb{P}\{W=w\}} \right].
\end{align*}
Therefore, we can rewrite the costs of the sender and the receiver, respectively, as
$U(\alpha,\beta)=\xi(\alpha,\beta)+\varrho \zeta(\alpha)$ and $V(\alpha,\beta)=\xi(\alpha,\beta)$.
Now, we can properly define the equilibrium of the  game.

\update{
\begin{definition} \label{def:equilibrium} (Nash Equilibrium): A pair $(\alpha^*,\beta^*)\in A\times B$ constitutes a Nash equilibrium of the privacy game if $(\alpha^*,\beta^*)\in\mathcal{N}$ with
\begin{align*}
\mathcal{N}=\{(\alpha,\beta)\in A\times B\,|\,&U(\alpha,\beta)\leq U(\alpha',\beta), \forall \alpha'\in A \\
&V(\alpha,\beta)\leq V(\alpha,\beta'), \forall \beta'\in B\}.
\end{align*}
\end{definition}
}

Now, we are ready to present the results of the paper.

\section{Finding an Equilibrium} \label{sec:results}
As all signalling games~\cite{JobelSignalling}, the privacy game admits a family of trivial equilibria known as \textit{babbling equilibria} in which the sender's message is independent of the to-be-estimated variable and the receiver  discards sender's message. 

\begin{theorem}(Babbling Equilibria): Let $\alpha^*\in A$ be such that $\alpha^*_{yzw}=1/|\mathcal{Y}|$ for all $(y,z,w)\in\mathcal{Y}\times\mathcal{X}\times \mathcal{W}$. Further, let $\beta^*\in B$ be such that $\beta^*_{\hat{x}y}=1$ for $\hat{x}\in \argmax_{x\in\mathcal{X}}\sum_{z'\in\mathcal{X}}\sum_{w'\in\mathcal{W}}p(x,w',z')$. Then, $(\alpha^*,\beta^*)$ constitutes an equilibrium.
\end{theorem}

\begin{IEEEproof} If the receiver does not use the transmitted message $Y$, the sender's best policy is to minimize $I(Y;W)$, which is achieved by employing a uniform distribution on $\mathcal{Y}$~\cite{Elements2006}. Furthermore, if the sender's message is independent of $(Z,W)$, the receiver's best policy is to set her estimate to be equal to the element with maximum \textit{ex ante} likelihood.\end{IEEEproof}

The messages passed at a babbling equilibrium are meaningless and do not contain any information. In what follows, we propose methods for capturing other equilibria of the game. To do so, we need to define a useful concept.

\begin{definition}(Potential Game~\cite{rosenthal1973class,monderer1996potential}): The privacy game admits a potential function $\Psi:A\times B\rightarrow \mathbb{R}$ if
\begin{align*}
V(\alpha,\beta)-V(\alpha,\beta')&=\Psi(\alpha,\beta)-\Psi(\alpha,\beta'),\\
U(\alpha,\beta)-U(\alpha',\beta)&=\Psi(\alpha,\beta)-\Psi(\alpha',\beta),
\end{align*}
for all $\alpha,\alpha'\in A$ and $\beta,\beta'\in B$. If the game admits a potential function, it is a potential game.
\end{definition}

The following simple, yet useful, lemma proves that the presented communication game admits a potential function.

\begin{lemma}[Potential Game] \label{lem:potential} The privacy game admits the potential function $\Psi(\alpha,\beta)=\xi(\alpha,\beta)+\varrho\zeta(\alpha)$. 
\end{lemma}

\begin{IEEEproof} First, note that, for the receiver, we have
\begin{align*}
V(\alpha,\beta)\hspace{-.03in}-\hspace{-.03in}V(\alpha,\beta')
&=\xi(\alpha,\beta)\hspace{-.03in}-\hspace{-.03in}\xi(\alpha,\beta')=\Psi(\alpha,\beta)\hspace{-.03in}-\hspace{-.03in}\Psi(\alpha,\beta'),
\end{align*}
and
\begin{align*}
U(\alpha,\beta)-U(\alpha',\beta)
&=\xi(\alpha,\beta)+\varrho\zeta(\alpha)-\xi(\alpha',\beta)-\varrho\zeta(\alpha')\\
&=\Psi(\alpha,\beta)-\Psi(\alpha',\beta).
\end{align*}
for all $\alpha,\alpha'\in A$ and $\beta,\beta'\in B$. 
\end{IEEEproof}

The result of Lemma~\ref{lem:potential} provides the following numerical method for constructing an equilibrium of the game. 

\begin{theorem} \label{tho:optim} Any $(\alpha^*,\beta^*)\in \argmin_{(\alpha,\beta)\in A\times B}\Psi(\alpha,\beta)$ constitutes an equilibrium of the game.
\end{theorem}

\begin{IEEEproof} The proof follows from Lemma~2.1 in~\cite{monderer1996potential}.
\end{IEEEproof}

Theorem~\ref{tho:optim} paves the way for constructing numerical methods to find an equilibrium of the game. This can be done by employing the various numerical optimization methods to minimize the potential function. However, this is a difficult task as the potential function is not convex (it is only convex in each variable separately and not in both variables simultaneously). We can simplify the construction of an equilibrium of the game for the special case where the transmitted message $Y$ and the to-be-estimated variable $X$ span over the same set.

\begin{algorithm}[t]
\caption{\label{alg:1} The best-response dynamics for learning an equilibrium.}
\begin{algorithmic}[1]
\REQUIRE $\alpha^0\in A$, $\beta^0\in B$
\FOR{$k=1,2,\dots$}
\IF{\update{$k$ is even}}
\STATE $\alpha^k\in \argmin_{\alpha\in A} U(\alpha,\beta^{k-1})$
\STATE $\beta^k \leftarrow\beta^{k-1}$
\ELSE
\STATE $\beta^k\in \argmin_{\beta\in B} V(\alpha^{k-1},\beta)$
\STATE $\alpha^k \leftarrow\alpha^{k-1}$
\ENDIF
\ENDFOR
\end{algorithmic}
\end{algorithm}

\begin{theorem} \label{tho:explicit} Assume that $\mathcal{Y}=\mathcal{X}$. Let $\beta'$ be such that $\beta'_{\hat{x}y}=1$ if $\hat{x}=y$ and $\beta'_{\hat{x}y}=0$ if $\hat{x}\neq y$. Moreover, let $\alpha'\in\argmin_{\alpha\in A} [\xi(\alpha,\beta')+\varrho\zeta(\alpha)]$. Then, $(\alpha',\beta')$ constitutes an equilibrium of the game. 
\end{theorem}

\begin{IEEEproof} Note that $\beta'$ means that $\hat{X}=Y$ with probability one, i.e., no data precessing is performed at the receiver. Clearly, if $\hat{X}=Y$, the sender finds $\alpha'$ so that $Y$ minimizes $\mathbb{E}\{d(X,Y)\}+\varrho I(W;Y).$ By definition, this is equivalent of saying that $\alpha'\in\argmin_{\alpha\in A} [\xi(\alpha,\beta')+\varrho\zeta(\alpha)]$. In the rest of the proof, we show that the best response of the receiver is to use $\beta'$. We do this by \textit{reductio ad absurdum}. Assume that there exists $\hat{X}$ constructed according to the conditional distribution $\mathbb{P}\{\hat{X}=\hat{x}|\mathcal{Y}=y\}=\beta_{\hat{x}y}$, for all $\hat{x},y\in\mathcal{X}$, such that $\mathbb{E}\{d(X,\hat{X})\}< \mathbb{E}\{d(X,Y)\}$ (because otherwise the receiver sticks to $\beta'$). Following the data processing inequality from Theorem 2.8.1~\cite[p.\,34]{Elements2006}, we have $I(W;\hat{X})\leq I(W;Y)$. This shows that 
$\mathbb{E}\{d(X,\hat{X})\}+\varrho I(W;\hat{X})<\mathbb{E}\{d(X,Y)\}+\varrho I(W;Y).$ This is evidently in contradiction with the optimality of $\alpha'$.\end{IEEEproof}

\begin{remark} \label{remark:duality} The proof of Theorem~\ref{tho:explicit} reveals that the sender's policy is the solution of the optimization problem
$\min_{\alpha\in A} \mathbb{E}\{d(X,Y)\}+\varrho I(W;Y).$
This problem is equivalent with solving $\min_{\alpha\in A:I(W;Y)\leq \vartheta}\mathbb{E}\{d(X,Y)\}$
where $\vartheta$ is an appropriate function of $\varrho$. Therefore, intuitively, the sender aims at providing an accurate measurement of the state $X$ while bounding the amount of the leaked information.
\end{remark}

For more general cases, we can use a distributed learning algorithm to recover an equilibrium. An example of such a learning algorithm is the iterative best-response dynamics. Following this, we can construct Algorithm~\ref{alg:1} to  recover an equilibrium of the game distributedly. \update{To present our results, we need to introduce a more practical notion of equilibrium. 

\begin{definition} \label{def:eps_equilibrium} ($\epsilon$-Nash Equilibrium): For all $\epsilon>0$, a pair $(\alpha^*,\beta^*)\in A\times B$ constitutes an $\epsilon$-Nash equilibrium of the privacy game if $(\alpha^*,\beta^*)\in\mathcal{N}_\epsilon$ with
\begin{align*}
\mathcal{N}_\epsilon=\{(\alpha,\beta)\in A\times B\,|\,&U(\alpha,\beta)\leq U(\alpha',\beta)+\epsilon, \forall \alpha'\in A \\
&V(\alpha,\beta)\leq V(\alpha,\beta')+\epsilon, \forall \beta'\in B\}.
\end{align*}
\end{definition}

This notion of equilibrium means that each player cannot gain by more than $\epsilon$ from unilaterally changing her actions, which is a practical notion if the act of changing her actions has ``some cost'' for the player. Now, we are ready to prove that Algorithm~\ref{alg:1} can extract an $\epsilon$-Nash equilibrium. 
}

\begin{theorem} \label{tho:convergence} 
\update{For $\{(\alpha^{k},\beta^{k})\}_{k\in\mathbb{N}}$ generated by Algorithm~\ref{alg:1} and all $\epsilon>0$, there exists $K_\epsilon\in\mathbb{N}$ such that $(\alpha^{k},\beta^{k})\in\mathcal{N}_\epsilon$ for all $k\geq K_\epsilon$.}
\end{theorem}

\ifdefined\longpaper
\begin{IEEEproof} \update{The proof is done by \textit{reductio ad absurdum}. To do so, assume that there exists an increasing subsequence $\{k_z\}_{z\in\mathbb{N}}$ such that $(\alpha^{k_z},\beta^{k_z})\notin\mathcal{N}_\epsilon,\forall z\in\mathbb{N}$.
If $(\alpha^{k},\beta^{k})\notin\mathcal{N}_\epsilon$ for some $k\geq 3$, at least one of the following cases hold.
\par $\bullet$ Case~1 ($\exists \alpha'\in A: U(\alpha^k,\beta^k)> U(\alpha',\beta^k)+\epsilon$ and $k$ is even): This means that $\beta^k=\beta^{k-1}$. Thus, we know that there exists $\alpha'\in A$ such that $U(\alpha',\beta^{k-1})<U(\alpha^k,\beta^{k-1})-\epsilon$. This is in contradiction with Line~3 of Algorithm~\ref{alg:1} and, thus, will never occur.
\par $\bullet$ Case~2 ($\exists \alpha'\in A: U(\alpha^k,\beta^k)> U(\alpha',\beta^k)+\epsilon$ and $k$ is odd): In this case, we have
\begin{align*}
\Psi(\alpha&^{k+1},\beta^{k+1})-\Psi(\alpha^{k},\beta^{k})\\
&=\Psi(\alpha^{k+1},\beta^{k})-\Psi(\alpha^{k},\beta^{k})\\
&=U(\alpha^{k+1},\beta^{k})-U(\alpha^{k},\beta^{k}) && k+1\mbox{ is even}\\
&\leq U(\alpha',\beta^{k})-U(\alpha^{k},\beta^{k}) && \mbox{Line~3 in Algorithm~\ref{alg:1}}\\
&<-\epsilon.
\end{align*}
\par $\bullet$ Case~3 ($\exists \beta'\in B: V(\alpha^k,\beta^k)> V(\alpha^k,\beta')+\epsilon$ and $k$ is even): In this case, we have
\begin{align*}
\Psi(\alpha&^{k+1},\beta^{k+1})-\Psi(\alpha^{k},\beta^{k})\\
&=\Psi(\alpha^{k},\beta^{k+1})-\Psi(\alpha^{k},\beta^{k})\\
&=V(\alpha^{k},\beta^{k+1})-V(\alpha^{k},\beta^{k}) && k+1\mbox{ is odd}\\
&\leq V(\alpha^{k},\beta')-V(\alpha^{k},\beta^{k}) && \mbox{Line~6 in Algorithm~\ref{alg:1}}\\
&<-\epsilon.
\end{align*}
\par $\bullet$ Case~4 ($\exists \beta'\in B: V(\alpha^k,\beta^k)> V(\alpha^k,\beta')+\epsilon$ and $k$ is odd): This means that $\alpha^k=\alpha^{k-1}$. Further, we know that there exists $\beta'\in A$ such that $V(\alpha^{k-1},\beta')<V(\alpha^{k-1},\beta^{k})-\epsilon$. This is in contradiction with Line~6 of Algorithm~\ref{alg:1} and, thus, will never occur.
\par From combining Cases~1--4, we know that if $(\alpha^{k},\beta^{k})\notin\mathcal{N}_\epsilon$ for some $k\geq 3$, then
$\Psi(\alpha^{k+1},\beta^{k+1})-\Psi(\alpha^{k},\beta^{k})<-\epsilon.$
Note that, in general, by construction of Line~3 in Algorithm~\ref{alg:1}, if $k$ is an even number, we get
\begin{align}
\Psi(\alpha^k,\beta^k)&-\Psi(\alpha^{k-1},\beta^{k-1})\nonumber\\
&=U(\alpha^k,\beta^{k-1})-U(\alpha^{k-1},\beta^{k-1})\leq 0.  \label{eqn:decreasing:even}
\end{align}
Similarly, by construction of Line~6 in Algorithm~\ref{alg:1}, if $k$ is an odd number, we have
\begin{align}
\Psi(\alpha^k,\beta^k)&-\Psi(\alpha^{k-1},\beta^{k-1})\nonumber\\&=V(\alpha^{k-1},\beta^k)-V(\alpha^{k-1},\beta^{k-1})\leq 0. \label{eqn:decreasing:odd}
\end{align}
Therefore, we can deduce that
\begin{align*}
\lim_{k\rightarrow\infty}\hspace{-.04in}\Psi(\alpha^{k},\beta^{k})
\hspace{-.03in}=&\Psi(\alpha^{0},\beta^{0})+\hspace{-.15in}\sum_{t\in\mathbb{N}\cup\{0\}}\hspace{-.1in} [\Psi(\alpha^{t+1},\beta^{t+1})\hspace{-.03in}-\hspace{-.03in}\Psi(\alpha^{t},\beta^{t})]\\
\leq & -\hspace{-.16in}\sum_{z\in\mathbb{N}:k_z\geq 3} \hspace{-.13in} [\Psi(\alpha^{k_z+1},\beta^{k_z+1})-\Psi(\alpha^{k_z},\beta^{k_z})]\\
=&-\infty,
\end{align*}
which is in contradiction with that $\lim_{k\rightarrow\infty}\Psi(\alpha^{k},\beta^{k})$ exists, because, by~\eqref{eqn:decreasing:even} and~\eqref{eqn:decreasing:odd}, $\{\Psi(\alpha^{k},\beta^{k})\}_{k\in\mathbb{N}}$ is a monotone decreasing sequence and is lower bounded by zero. 
}
\end{IEEEproof}
\fi

\ifdefined\shortpaper
\begin{IEEEproof}
See~\cite{farokhi_long_note} for a detailed proof.
\end{IEEEproof}
\fi
\update{Theorem~\ref{tho:convergence} shows that Algorithm~\ref{alg:1} converges to an $\epsilon$-Nash equilibrium, for any $\epsilon>0$, in a finite number of iterations. We can slightly tweak Algorithm~\ref{alg:1} to also present bounds on the required number of iterations to extract an  $\epsilon$-Nash equilibrium.

\begin{algorithm}[t]
\caption{\label{alg:1_updated} The best-response dynamics for learning an equilibrium.}
\begin{algorithmic}[1]
\REQUIRE $\alpha^0\in A$, $\beta^0\in B$
\FOR{$k=1,2,\dots$}
\IF{\update{$k$ is even}} \update{
\STATE $\alpha'\in \argmin_{\alpha\in A} U(\alpha,\beta^{k-1})$
\IF{$U(\alpha^{k-1},\beta^{k-1})-U(\alpha',\beta^{k-1})>\epsilon$}
\STATE $\alpha^{k}\leftarrow \alpha'$
\ELSE
\STATE $\alpha^{k}\leftarrow \alpha^{k-1}$
\ENDIF}
\STATE $\beta^k \leftarrow\beta^{k-1}$
\ELSE
\STATE \update{$\beta^k\in \argmin_{\beta\in B} V(\alpha^{k-1},\beta)$
\IF{$V(\alpha^{k-1},\beta^{k-1})-V(\alpha^{k-1},\beta')>\epsilon$}
\STATE $\beta^{k}\leftarrow \beta'$
\ELSE
\STATE $\beta^{k}\leftarrow \beta^{k-1}$
\ENDIF}
\STATE $\alpha^k \leftarrow\alpha^{k-1}$
\ENDIF
\ENDFOR
\end{algorithmic}
\end{algorithm}

\begin{theorem} \label{tho:convergence_time} For $\{(\alpha^{k},\beta^{k})\}_{k\in\mathbb{N}}$ generated by Algorithm~\ref{alg:1_updated} and all $\epsilon>0$, $(\alpha^{k},\beta^{k})\in\mathcal{N}_\epsilon$ for all $k\geq 3+\Psi(\alpha^0,\beta^0)/\epsilon$.
\end{theorem}

\ifdefined\longpaper
\begin{IEEEproof} Algorithm~\ref{alg:1_updated} makes sure that $(\alpha^{k},\beta^{k})=(\alpha^{k-1},\beta^{k-1})$ if $(\alpha^{k},\beta^{k})\in\mathcal{N}_\epsilon$. Therefore, there exists $K$ such that $(\alpha^{k},\beta^{k})\notin\mathcal{N}_\epsilon$ for $k\leq K-1$ and $(\alpha^{k},\beta^{k})\in\mathcal{N}_\epsilon$ for $k\geq K$. Now, following the reasoning of the proof of Theorem~\ref{tho:convergence}, we can see that
\begin{align*}
\Psi(\alpha^{K},\beta^{K})
=&\Psi(\alpha^{2},\beta^{2})+\sum_{t=2}^{K-1} [\Psi(\alpha^{t+1},\beta^{t+1})-\Psi(\alpha^{t},\beta^{t})]\\
\leq & \Psi(\alpha^{2},\beta^{2})\hspace{-.03in}-\hspace{-.03in}(K-3)\epsilon
\hspace{-.03in}\leq \hspace{-.03in} \Psi(\alpha^0,\beta^0)-(K-3)\epsilon,
\end{align*}
where the last inequality follows from that $\Psi(\alpha^{k},\beta^{k})$ is a decreasing sequence. Noting that $\Psi(\alpha^{K},\beta^{K})\geq 0$ because of the properties of the mutual information and expected estimation error, we can see that $K\leq 3+\Psi(\alpha^0,\beta^0)/\epsilon$.
\end{IEEEproof}
\fi

\ifdefined\shortpaper
\begin{IEEEproof}
See~\cite{farokhi_long_note} for a detailed proof.
\end{IEEEproof}
\fi

}

\newcommand{\n}{\mathbf{N}}
\section{Extension to Multiple Senders} \label{sec:multiple}
Here, we extend the results to the case where sender $S_i$, $1\leq i\leq n$, for some $n\geq 2$, communicate with the receiver $R$. Similarly, the receiver wants to have an accurate measurement of a random variable $X\in\mathcal{X}$. We assume that sender $i$ has access to a possibly noisy measurement of the state denoted by $Z_i\in\mathcal{X}$. The private information of the senders is denoted by $W_i\in\mathcal{W}_i$. The senders would like to transmit a message $Y_i\in\mathcal{Y}_i$ that contains useful information about the measured state $Z_i$ while minimizing the amount of the leaked private information. Let  $\n=\{1,\dots,n\}$. 

\begin{assumption} The random variables $X,(Z_i)_{i\in\n},(W_i)_{i\in\n}$ are distributed according to a joint probability distribution $p:\mathcal{X}\times\mathcal{X}^n\times\prod_{i\in\n}\mathcal{W}_i\rightarrow [0,1]$, i.e., $\mathbb{P}\{X=x,(Z_i)_{i\in\n}=(z_i)_{i\in\n},(W_i)_{i\in\n}=(w_i)_{i\in\n}\}=p(x,(z_i)_{i\in\n},(w_i)_{i\in\n})$ for all $(x,(z_i)_{i\in\n},(w_i)_{i\in\n})\in\mathcal{X}\times\mathcal{X}^n\times \prod_{i\in\n}\mathcal{W}_i$.
\end{assumption}

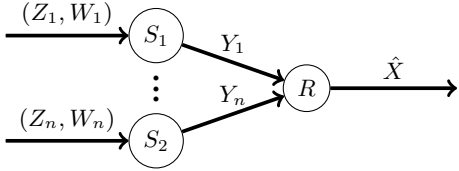
\begin{figure}[t]
\centering
\begin{tikzpicture}
\node[state,minimum size=0.1cm,scale=0.9] (R)  at (+1.0,+0.0) {$R$};
\node[state,minimum size=0.1cm,scale=0.9] (S1)  at (-1.0,+0.7) {$S_1$};
\node[state,minimum size=0.1cm,scale=0.9] (S2)  at (-1.0,-0.7) {$S_2$};
\node[] ()  at (-1.0,+0.1) {\LARGE $\vdots$};
\path[every node/.style={font=\sffamily\small}]
(S1) edge[->,double=black] node[above]{$Y_1$} (R)
(S2) edge[->,double=black] node[above]{$Y_n$} (R)
(R) edge[->,double=black] node[above]{$\hat{X}$} (+3.0,+0.0)
(-3.0,+0.7) edge[->,double=black] node[above] {$(Z_1,W_1)$} (S1)
(-3.0,-0.7) edge[->,double=black] node[above] {$(Z_n,W_n)$} (S2);
\end{tikzpicture}
\vspace{-.1in}
\caption{\label{fig:diagram1} Communication structure between the senders $S_1,\dots,S_n$ and the receiver $R$.}
\vspace{-.2in}
\end{figure}

\begin{algorithm}[t]
\caption{\label{alg:2} The best-response dynamics for learning an equilibrium in the multi-sensor case.}
\begin{algorithmic}[1]
\REQUIRE $\alpha^{(i),0}\in A$ for all $i\in\n$, $\beta^0\in B$
\FOR{$k=1,2,\dots$}
\STATE Select $j\in\{0,1,\dots,n\}$ uniformly at random
\IF{$j=0$}
\STATE $\beta^k\in \argmin_{\beta\in B'} V(\beta,(\alpha^{(i),k-1})_{i\in\n})$
\STATE $\alpha^{(i),k} \leftarrow\alpha^{(i),k-1}$ for all $i\in\n$
\ELSE
\STATE $\alpha^{(j),k}\in \argmin_{\alpha^{(j)}} U_j(\alpha^{(j)},(\alpha^{(i),k-1})_{j\neq i},\beta^{k-1})$ 
\STATE $\alpha^{(i),k} \leftarrow \alpha^{(i),k-1}$ for all $i\in\n\setminus\{j\}$
\STATE $\beta^k \leftarrow\beta^{k-1}$
\ENDIF
\ENDFOR
\end{algorithmic}
\end{algorithm}

Similarly, the receiver constructs its best estimate $\hat{X}\in\mathcal{X}$ using the conditional distribution $\mathbb{P}\{\hat{X}=\hat{x}|(Y_i)_{i\in\n}=(y_i)_{i\in\n}\}=\beta_{\hat{x}y_1\dots y_n}$ for all $(\hat{x},(y_i)_{i\in\n})\in\mathcal{X}\times\prod_{i\in\n}\mathcal{Y}_i$. The tensor $\beta=(\beta_{\hat{x}y_1\dots y_n})_{(\hat{x},(y_i)_{i\in\n})\in\mathcal{X}\times\prod_{i\in\n}\mathcal{Y}_i}\in B'$ is the policy of the receiver with the set of feasible policies defined~as
\begin{align*}
B'=\bigg\{\beta:\,&\beta_{\hat{x}y_1\dots y_n}\in[0,1],\forall (\hat{x},(y_i)_{i\in\n})\in\mathcal{X}\times\prod_{i\in\n}\mathcal{Y}_i\\[-.4em] &\mbox{ and }\sum_{\hat{x}\in\mathcal{X}}\beta_{\hat{x}y_1\dots y_n}=1,\forall (y_i)_{i\in\n}\in\prod_{i\in\n}\mathcal{Y}_i\bigg\}.
\end{align*}
The receiver seeks an accurate measurement of the variable $X$ and, hence, minimizes the cost function 
$\mathbb{E}\{d(X,\hat{X})\}$.

The sender $i\in\n$ constructs its message $y_i\in\mathcal{Y}_i$ according to the conditional probability distribution $\mathbb{P}\{Y=y_i|Z=z_i,W=w_i\}=\alpha^{(i)}_{y_iz_iw_i}$ for all $(y_i,z_i,w_i)\in\mathcal{Y}_i\times\mathcal{X}\times \mathcal{W}_i$. The tensor $\alpha^{(i)}=(\alpha^{(i)}_{y_iz_iw_i})_{(y_i,z_i,w_i)\in\mathcal{Y}_i\times\mathcal{X}\times \mathcal{W}_i}\in A'_i$ denotes the policy of the sender. The set of feasible policies is 
\begin{align*}
A'_i=\bigg\{&\alpha^{(i)}: \alpha^{(i)}_{y_iz_iw_i}\in[0,1],\forall (y_i,z_i,w_i)\in\mathcal{Y}_i\times\mathcal{X}\times \mathcal{W}_i \\[-.4em]&\mbox{ and }\sum_{y_i\in\mathcal{Y}_i}\alpha^{(i)}_{y_iz_iw_i}=1,\forall (z_i,w_i)\in\mathcal{X}\times \mathcal{W}_i\bigg\}.
\end{align*}
The sender wants to minimize $\mathbb{E}\{d(X,\hat{X})\}+\varrho I(Y_i;W_i)$ to find a balance between transmitting useful information about the measured variable and maintaining her privacy. 

\begin{remark} This formulation is useful when $X$ and $W_i$ are uncorrelated while $Z_i$ and $W_i$ are correlated (e.g., participatory-sensing mechanisms for traffic estimation). This is because, in this formulation, the sender is only concerned about the amount of the leaked information in her own message $I(Y_i;W_i)$. However, if $X$ and $W_i$ were to be  correlated, she should have been concerned with the total amount of the leaked information $I(Y_1,\dots,Y_n;W_i)$. This is indeed the case because the receiver can construct an accurate representation of the state (if each sensor is only concerned about the amount of the leaked information in her own message) and, therefore, gain an insight into the private information of the sensors.
\end{remark}

Following similar calculations as in Section~\ref{sec:results}, the cost of the receiver is given by
\begin{align*}
V((\alpha^{(i)})_{i\in\n},\beta)
&=\mathbb{E}\{d(X,\hat{X})\}=\xi'(\beta,(\alpha^{(i)})_{i\in\n}),
\end{align*}
where the mapping $\xi':B'\times\prod_{i\in\n}A'_i\rightarrow [0,1]$ is defined as in~\eqref{eqn:long:multiplayer}, on top of the next page, for all $(\beta,(\alpha^{(i)})_{i\in\n})\times B'\times\prod_{i\in\n}A'_i$.
\begin{figure*}
\begin{align} 
\xi'(\beta,(\alpha^{(i)})_{i\in\n})
=\hspace{-.2in}\sum_{(y_1,z_1,w_1)\in\mathcal{Y}_1\times\mathcal{X}\times \mathcal{W}_1}\cdots \sum_{(y_n,z_n,w_n)\in\mathcal{Y}_n\times\mathcal{X}\times \mathcal{W}_n}\sum_{\hat{x},x\in\mathcal{X}}
d(x,\hat{x})\beta_{\hat{x}y_1\dots y_n}\prod_{i\in\n}\alpha^{(i)}_{y_iz_iw_i}p(x,(z_i)_{i\in\n},(w_i)_{i\in\n})\label{eqn:long:multiplayer}
\end{align}
\hrule
\vspace{-.2in}
\end{figure*}
The cost of sender $j\in\n$ is given by
\begin{align*}
U_j((\alpha^{(i)})_{i\in\n},\beta)
&=\mathbb{E}\{d(X,\hat{X})\}+\varrho I(Y_j;W_j)\\
&=\xi'(\beta,(\alpha^{(i)})_{i\in\n})+\varrho\zeta'_j(\alpha^{(j)}),
\end{align*}
where 
\begin{align*}
\zeta'_j(\alpha^{(j)})=\sum_{y_j\in\mathcal{Y}_j}&\sum_{w_j\in\mathcal{W}_j} \mathbb{P}\{Y_j=y_j,W_j=w_j\} \nonumber \\& \times \log\left[\frac{\mathbb{P}\{Y_j=y_j,W_j=w_j\}}{\mathbb{P}\{Y_j=y_j\}\mathbb{P}\{W_j=w_j\}} \right].
\end{align*}

\update{
\begin{definition}(Nash Equilibrium): A pair $(\hspace{-.02in}(\hspace{-.02in}\alpha^{\hspace{-.02in}(i)\hspace{-.01in},\hspace{-.02in}*})_{i\in\n}\hspace{-.02in},\hspace{-.02in}\beta^*\hspace{-.02in})\in \prod_{i\in\n}A'_i\times B'$ constitutes a Nash equilibrium of the privacy game if $((\alpha^{(i),*})_{i\in\n},\beta^*)\in\mathcal{N}'$ with
\begin{align*}
\mathcal{N}'=\bigg\{&((\alpha^{(i)})_{i\in\n},\beta)\in \prod_{i\in\n}A'_i\times B'\,|\,\forall \bar{\alpha}^{(j)}\in A'_j,\forall j\in\n,\\
&U_j((\alpha^{(i)})_{i\in\n},\beta)\leq U_j(\bar{\alpha}^{(j)},(\alpha^{(i)})_{i\in\n\setminus\{j\}},\beta), \\
&V((\alpha^{(i)})_{i\in\n},\beta)\leq V((\alpha^{(i)})_{i\in\n},\bar{\beta}), \forall \bar{\beta}\in B'\bigg\}.
\end{align*}
\end{definition}

\begin{definition}(Potential Game): The defined game admits a potential function $\Psi':\prod_{i\in\n}A'_i\times B'\rightarrow \mathbb{R}$ if
\begin{align*}
V((\alpha^{(i)}&)_{i\in\n},\beta)-V((\alpha^{(i)})_{i\in\n},\bar{\beta})\\
&=\Psi'((\alpha^{(i)})_{i\in\n},\beta)-\Psi((\alpha^{(i)})_{i\in\n},\bar{\beta}),\\
U_j((\alpha^{(i)}&)_{i\in\n},\beta)-U_j(\bar{\alpha}^{(j)},(\alpha^{(i)})_{i\in\n\setminus\{j\}},\beta)\\
&=\Psi'((\alpha^{(i)})_{i\in\n},\beta)-\Psi(\bar{\alpha}^{(j)},(\alpha^{(i)})_{i\in\n\setminus\{j\}},\beta),
\end{align*}
for all $((\alpha^{(i)})_{i\in\n},\beta)\in\prod_{i\in\n}A'_i\times B'$, $\bar{\alpha}^{(j)}\in A'_j$, $j\in\n$, and $\bar{\beta}\in B'$. If the game admits a potential function, it is a potential game.
\end{definition}
}

\begin{lemma} \label{lem:multi:potential} The game admits the potential function $\Psi'((\alpha^{(i)})_{i\in\n},\beta)=\xi'(\beta,(\alpha^{(i)})_{i\in\n})+\varrho\sum_{i\in\n}\zeta_i(\alpha^{(i)})$.
\end{lemma}

\begin{IEEEproof} The proof of this lemma is, \textit{mutatis mutandis}, similar to that of Lemma~\ref{lem:potential}.
\end{IEEEproof}

\begin{theorem} Any $((\alpha^{(i)*})_{i\in\n},\beta^*)$ in $\argmin_{((\alpha^{(i)})_{i\in\n},\beta)\in \prod_{i\in\n}A'_i\times B'}\Psi'((\alpha^{(i)})_{i\in\n},\beta)
$
constitutes an equilibrium of the game.
\end{theorem}

\begin{IEEEproof} The proof follows from Lemma~2.1 in~\cite{monderer1996potential}.
\end{IEEEproof}

\update{\begin{definition} \label{def:eps_equilibrium} ($\epsilon$-Nash Equilibrium): For all $\epsilon>0$, a pair $((\alpha^{(i),*})_{i\in\n},\beta^*)\in \prod_{i\in\n}A'_i\times B'$ constitutes an $\epsilon$-Nash equilibrium of the privacy game if $((\alpha^{(i),*})_{i\in\n},\beta^*)\in\mathcal{N}'_\epsilon$ with
\begin{align*}
\mathcal{N}'_\epsilon=\bigg\{&((\alpha^{(i)})_{i\in\n},\beta)\in \prod_{i\in\n}A'_i\times B'\,|\,\forall \bar{\alpha}^{(j)}\in A'_j,\forall j\in\n,\\
&U_j((\alpha^{(i)})_{i\in\n},\beta)\hspace{-.04in}\leq\hspace{-.04in} U_j(\bar{\alpha}^{(j)},(\alpha^{(i)})_{i\in\n\setminus\{j\}},\beta)\hspace{-.04in}+\hspace{-.04in}\epsilon, \\
&V((\alpha^{(i)})_{i\in\n},\beta)\leq V((\alpha^{(i)})_{i\in\n},\bar{\beta})+\epsilon, \forall \bar{\beta}\in B\bigg\}.
\end{align*}
\end{definition}}

In the multi-sensor case, we can use Algorithm~\ref{alg:2} to recover an equilibrium of the game using an iterative best-response dynamics.

\begin{theorem} \label{tho:convergence_multiple}
\update{ For $\{((\alpha^{(i),k})_{i\in\n},\beta^{k})\}_{k\in\mathbb{N}}$ generated by Algorithm~\ref{alg:2} and all $\epsilon>0$,}
$\lim_{k\rightarrow \infty} \mathbb{P}\{((\alpha^{(i),k})_{i\in\n},\beta^{k})\hspace{-.03in}\in\hspace{-.03in}\mathcal{N}_\epsilon\}=1$.
\end{theorem}

\ifdefined\longpaper
\begin{IEEEproof} 
\update{The proof is done by \textit{reductio ad absurdum}. To do so, assume that there exists an increasing subsequence $\{k_z\}_{z\in\mathbb{N}}$ and a constant $\delta>0$ such that $\mathbb{P}\{(\alpha^{k_z},\beta^{k_z})\notin\mathcal{N}'_\epsilon\}>\delta$ for all $z\in\mathbb{N}$. 
First, note that, with a similar reasoning as in the proof of Theorem~\ref{tho:convergence}, we can show that, for all $k\in\mathbb{N}$,
$\Psi'((\alpha^{(i),k+1})_{i\in\n},\beta^{k+1})-\Psi'((\alpha^{(i),k})_{i\in\n},\beta^{k})\leq 0$.
Now, if $(\alpha^k,\beta^k)\notin\mathcal{N}'_\epsilon$ then, either $\exists\ell\in\n,\exists \bar{\alpha}^{\ell}\in A'_\ell: U_\ell((\alpha^{(i),k})_{i\in\n},\beta^{k})> U_\ell(\bar{\alpha}^{(\ell)},(\alpha^{(i),k})_{i\in\n\setminus\{\ell\}},\beta^{k})+\epsilon$ for some $\ell\in\n$ or $\exists \bar{\beta}\in B': V((\alpha^{(i),k})_{i\in\n},\beta^{k})> V((\alpha^{(i),k})_{i\in\n},\bar{\beta})+\epsilon$. If $\exists \bar{\beta}\in B': V((\alpha^{(i),k})_{i\in\n},\beta^{k})> V((\alpha^{(i),k})_{i\in\n},\bar{\beta})+\epsilon$ and if $j=0$, we get that $\Psi'((\alpha^{(i),k+1})_{i\in\n},\beta^{k+1})-\Psi'((\alpha^{(i),k})_{i\in\n},\beta^{k}) <-\epsilon.$ Therefore, if $\exists \bar{\beta}\in B': V((\alpha^{(i),k})_{i\in\n},\beta^{k})> V((\alpha^{(i),k})_{i\in\n},\bar{\beta})+\epsilon$, the inequality in~\eqref{eqn:long:proof:1}, on top of the next page, holds. 
\begin{figure*}
\begin{align} \label{eqn:long:proof:1}
\mathbb{E}\{\Psi'((\alpha^{(i),k+1})_{i\in\n},\beta^{k+1})\hspace{-.03in}-\hspace{-.03in}\Psi'((\alpha^{(i),k})_{i\in\n},\beta^{k})\}
&\hspace{-.03in}=\hspace{-.03in}\sum_{q=0}^n \mathbb{E}\{\Psi'((\alpha^{(i),k+1})_{i\in\n},\beta^{k+1})\hspace{-.03in}-\hspace{-.03in}\Psi'((\alpha^{(i),k})_{i\in\n},\beta^{k})|j=q\} \mathbb{P}\{j=q\}\nonumber\\
&<-\epsilon/(n+1).
\end{align}
\hrule
\vspace*{-.2in}
\end{figure*}
Similarly, if $\exists \bar{\alpha}^{\ell}\in A'_\ell: U_\ell((\alpha^{(i),k})_{i\in\n},\beta^{k})> U_\ell(\bar{\alpha}^{(\ell)},(\alpha^{(i),k})_{i\in\n\setminus\{\ell\}},\beta^{k})+\epsilon$ and if $j=\ell$, we get that $\Psi'((\alpha^{(i),k+1})_{i\in\n},\beta^{k+1})-\Psi'((\alpha^{(i),k})_{i\in\n},\beta^{k}) <-\epsilon$, which leads to the inequality~\eqref{eqn:long:proof:1}. Notice that
\begin{align*}
\lim_{k\rightarrow\infty}&\mathbb{E}\{\Psi((\alpha^{(i),k})_{i\in\n},\beta^{k})\}\\
=&\mathbb{E}\{\Psi((\alpha^{(i),0})_{i\in\n},\beta^{0})\}\\
&+\hspace{-.03in}\sum_{t=0}^\infty \mathbb{E}\{\Psi((\alpha^{(i),t+1})_{i\in\n},\beta^{t+1})\hspace{-.03in}-\hspace{-.03in}\Psi((\alpha^{(i),t})_{i\in\n},\beta^{t})\}\\
\leq & \mathbb{E}\{\Psi((\alpha^{(i),0})_{i\in\n},\beta^{0})\}\\
&-\epsilon/(n+1)\sum_{z=0}^\infty\mathbb{P}\{((\alpha^{(i),k_z})_{i\in\n},\beta^{k_z})\notin\mathcal{N}'_\epsilon\} \\
=&-\infty.
\end{align*}
This is in contradiction with that $\lim_{k\rightarrow\infty}\mathbb{E}\{\Psi(\alpha^{k},\beta^{k})\}$ exists and is greater than or equal to zero. Thus, we have proved that $\lim_{k\rightarrow\infty}\mathbb{P}\{(\alpha^k,\beta^k)\notin\mathcal{N}'_\epsilon\}=0$.
}
\end{IEEEproof}
\fi

\ifdefined\shortpaper
\begin{IEEEproof}
See~\cite{farokhi_long_note} for a detailed proof.
\end{IEEEproof}
\fi

\section{Numerical Example} \label{sec:numerical}
Consider an example with $\mathcal{X}=\mathcal{W}=\mathcal{Y}=\{1,\dots,5\}$. Assume that $Z=X$, i.e., the sender has access to the perfect measurement of $X$. Moreover, let
\begin{align*}
(\mathbb{P}\{X=x,W=&w\})_{x\in\mathcal{X},w\in\mathcal{W}}\\
&=\matrix{ccccc}{
0.14 & 0.02 & 0.01 & 0.01 & 0.02 \\
0.02 & 0.14 & 0.02 & 0.01 & 0.01 \\
0.01 & 0.02 & 0.14 & 0.02 & 0.01 \\
0.01 & 0.01 & 0.02 & 0.14 & 0.02 \\
0.02 & 0.01 & 0.01 & 0.02 & 0.14
}.
\end{align*}
This distribution implies that there is a reasonable correlation between $X$ and $W$. Therefore, for high enough $\varrho$, we expect a very bad estimation quality (at the receiver) since, otherwise, the receiver can recover $X$, which carries a significant amount of information about $W$.

\begin{figure}
\centering
\begin{tikzpicture}[>=stealth']
\node[] at (0,0) {\includegraphics[width=1.0\linewidth]{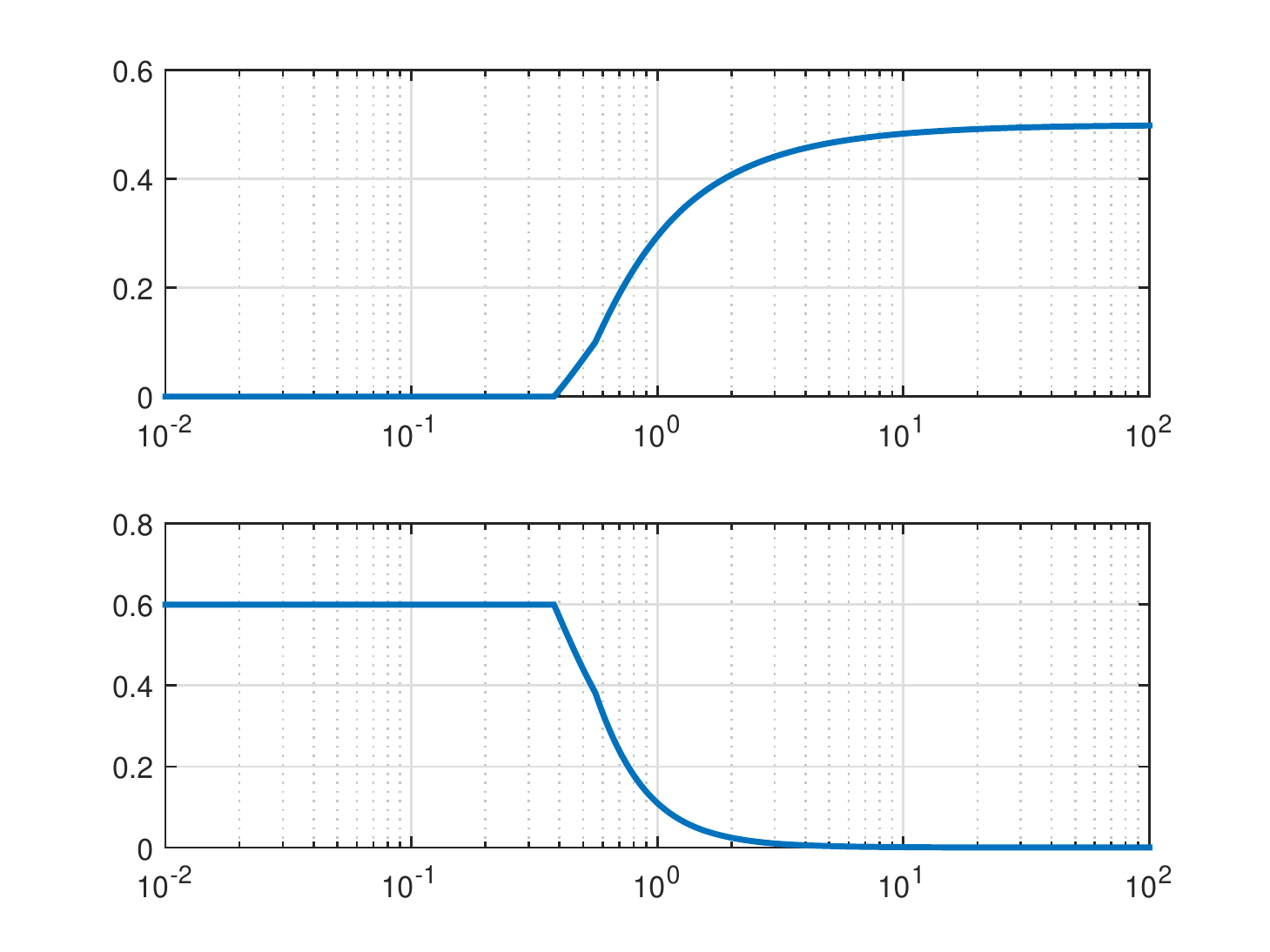}};
\node[fill=white] at (-1.8,1.8) {\includegraphics[width=0.28\linewidth]{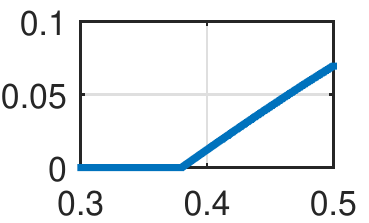}};
\node[rotate=90] at (-4.0,1.8) {\small $\mathbb{E}\{d(X,\hat{X})\}$};
\node[rotate=90] at (-4.0,-1.5) {\small $I(Y;W)$};
\draw[thick] (-0.73,+0.56) -- 
			 (-0.73,+0.95) -- 
			 (-0.35,+0.95) -- 
			 (-0.35,+0.56) -- 
			 (-0.73,+0.56);
\path[->] (-0.73,+0.72) edge[] (-1.22,+1.2);
\node[] at (0.2,-3.2) {$\varrho$};
\end{tikzpicture} \vspace{-.3in}
\caption{\label{figure:numerical} Expected estimation error $\mathbb{E}\{d(X,\hat{X})\}$ and mutual information $I(Y;W)$ at the extracted equilibrium versus the privacy ratio $\varrho$. }
\vspace{-.2in}
\end{figure}
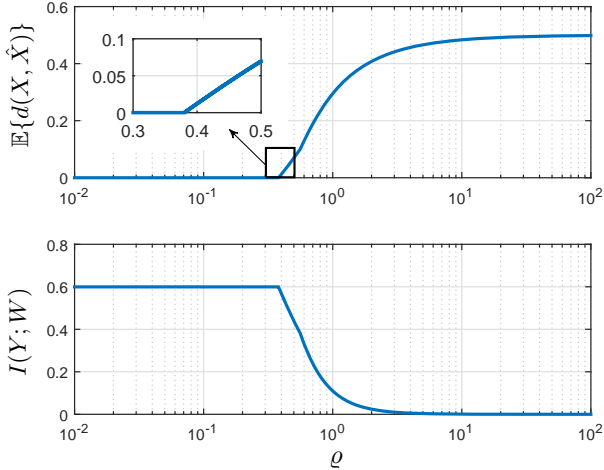

For this example, we use the results of Theorem~\ref{tho:explicit} extract a nontrivial equilibrium for each $\varrho$. Fig.~\ref{figure:numerical}~(top) illustrate the estimation error $\mathbb{E}\{d(X,\hat{X})\}$ as a function of the privacy ratio $\varrho$. As we expect, by increasing $\varrho$, the sender puts more emphasis on protecting her privacy rather than providing a useful measurement to the receiver and, therefore, $\mathbb{E}\{d(X,\hat{X})\}$ increases. Fig.~\ref{figure:numerical}~(bottom) shows the mutual information $I(Y;W)$ as a function of the privacy ratio $\varrho$. Evidently, with increasing $\varrho$, the amount of leaked information about the private information of the sender decreases. In both figures, there seems to be sudden change when the privacy ratio passes the critical value $\varrho=0.38$. That is, for all $\varrho<0.38$, the truth-telling seems to be an equilibrium of the game; however, if $\varrho>0.38$, the sender adds false reports to protects her privacy.

\section{Conclusion} \label{sec:conclusions}
We developed a game-theoretic framework to investigate the effect of privacy in the quality of the measurements provided by a well-informed sender to a receiver. We used a privacy ratio to model the sender's emphasis on protecting her privacy. Equilibria of the game were constructed. We proposed learning algorithms for recovering an equilibrium.  Future work can focus on extending the results to dynamic estimation problems.

\bibliographystyle{IEEEtran}
\bibliography{citation}

\end{document}